\documentclass[aps,prd,preprint,groupedaddress,showpacs]{revtex4}
\usepackage{dcolumn}

\usepackage{bm}

\usepackage{graphicx}
\usepackage{subfigure}
\usepackage{verbatim}
\usepackage{amsmath}
\usepackage{amssymb}
\usepackage{mathrsfs}
\usepackage{booktabs}
\usepackage{slashed}

\begin{document}

%\hypersetup{colorlinks=true,
%citecolor=\textcolor[rgb]{0.00,0.00,1.00}{blue}}

%\preprint{}

\title{$0^{++}$ scalar glueball in finite-width Gaussian sum rules}

\author{Shuiguo Wen}
%\email{airywsg@163.com}
\author{Zhenyu Zhang}
%\email{jelime_d@163.com}
\author{Jueping Liu
\footnote{To whom correspondence should be addressed. Email:
jpliu@whu.edu.cn}}
\affiliation{%
College of Physics and Technology, Wuhan University, 430072 Wuhan,
China}


%\author{}
%\thanks{}
\affiliation{}


\date{\today}

\begin{abstract}
Based on a semiclassical expansion for quantum chromodynamics in
the instanton liquid background, the correlation function of the
$0^{++}$ scalar glueball current is given, and the properties of the
$0^{++}$ scalar glueball are studied in the framework of Gaussian
sum rules. Besides the pure classical and quantum contributions, the
contributions arising from the interactions between the classical
instanton fields and quantum gluons are come into play. Instead of
the usual zero-width approximation for the resonance, the
Breit-Wigner form for the spectral function of the finite-width
resonance is adopted. The family of the Gaussian sum rules for the
scalar glueball in quantum chromodynamics with and without light
quarks is studied. A consistency between the subtracted and
unsubtracted sum rules is very well justified, and the values of
the decay width and the coupling to the corresponding current for
the $0^{++}$ resonance, in which the scalar glueball fraction is
dominant, are obtained.
\end{abstract}

%\keywords{}

\maketitle


\section{Introduction}\label{sec:introduction}
Glueballs, being composed of pure gluons in the limit of no quark
fields, have attracted much attention since the theory of the strong
interactions, quantum chromodynamics (QCD), was founded in the late
of 1970s \cite{FGL73,FMi75,NSVZ80,NSVZ81}. Estimates of glueball
properties were obtained in a variety of approaches \cite{Mathieu09}, ranging from
model analyses
\cite{Albala08,Cheng06,Vicente06,Frank05,Aniso97,Szcze96,Michael05,Isgur85}
,quenched lattice QCD
(QLQCD)\cite{Meng09,Chen06,Colin99,Vaccar99,Sexton95} and unquenched
lattice QCD (UQLQCD) simulations \cite{Apoorva86,Chin86} to QCD sum
rule calculations
\cite{SVZ79,KJ01,Sch95,Forkel01,Forkel05,HSE01,HaS01,ZhS03,Narison98,JPLiu91,JPLiu93,Badesd89,Doming86}.

The lowest $0^{++}$ scalar glueball state is, in fact, the most
intricate hadron state which is difficult to figure out. In
lattice QCD (LQCD) the mass scale of the scalar glueball is predicted to
be in the rang of 1.3-1.7 GeV
\cite{Meng09,Chen06,Colin99,Vaccar99,Sexton95}. The
 $0^{++}$ scalar resonances closest to this energy range are
$f_0(1500)$ and $f_0(1710)$ in the present data \cite{Amsler08},
and some authors favor the former as the lightest scalar glueball
\cite{Amsler95,Amsler195}, while some others prefer the latter
\cite{Sexton95,Michael05}. Furthermore, both $f_0(1500)$ and
$f_0(1710)$ may not be pure glueballs; to the contrary, these
resonances can be considered to be the mixture of glueball and
mesons \cite{Cheng06,Vicente06,Frank05,Crede09}. In the QCD sum rule approach,
the results of the mass of the scalar glueball are also different
from each other. In the early days of QCD, some found a light
$0^{++}$ scalar glueball in the region of $300$-$700$ MeV
\cite{NSVZ80, JPLiu91,JPLiu93,Badesd89,Doming86} using the
subtracted sum rule (SSR), while the others obtained a much heavier
one in the $1$-$2$ GeV region \cite{Shifman81,Pascual82} by using
the unsubtracted sum rule (USSR). The inconsistency between both
subtracted and unsubtracted sum rules has bothered scientists for
many years.

It should be noticed that, in the early QCD sum rule approach, it
was already recognized that there is an onset of the nonperturbative
physics (a departure from the asymptotic freedom) in the scalar
glueball correlator at unusually short distances $|x|\ll
\Lambda^{-1}_{\mbox{QCD}}$ \cite{NSVZ81}. Such hard nonperturbative
effects are usually considered to be coming from direct instantons,
i.e. the tunneling processes which rearrange the QCD vacuum topology
in localized regions\cite{SS98,Dia03}. This physics was ignored in
the early glueball sum rules except in Ref. \cite{Shu82}.

Recently, a sizeable instanton contribution to the QCD sum rules of
the $0^{++}$ scalar glueball has been found on the basis of the
instanton liquid model of the QCD vacuum \cite{SS98,Dia03}, and
supported by the lattice simulations
\cite{Shu82,Perez99,Smith98,Hasenfratz98,Ringwald99}. However, the
instanton-induced continuum contributions are neglected in the early
QCD sum rule approach, and thus a reliable estimate of the glueball
properties cannot be obtained. On the basis of the
instanton-improved operator-product expansion (OPE), the authors in Refs. \cite{Forkel01,Sch95,
Forkel05} included the instanton-induced contributions to the
continuum spectrum, made a great improvement to the consistency
between different types of $0^{++}$ scalar glueball sum rules, and
gave new predictions for the mass and decay constant confirmed
later by the Gaussian sum rule (GSR) calculation \cite{HSE01}.
Moreover, we have clarified that the stability and the consistency
for the SSR and USSRs for the $0^{++}$ scalar glueball can be
obtained by using a systematical semiclassical expansion of the
background instanton fields in the instanton liquid model of the QCD
vacuum \cite{ZhangZY06}.

Instantons, which make a great difference in the scalar and
pseudoscalar channels \cite{GI80,NSVZ81,Shuryak83}, play a major
role in the gluonic interactions, especially in the nonperturbative
region \cite{SS98}. Direct instanton contributions are included in
the QCD sum rules \cite{KJ01,Sch95,Forkel01,Forkel05,HSE01}. The
results are found to be in good agreement with different USSRs. The
compatibility between the SSR and USSRs become much better but are not
satisfying yet, because only the leading classic effects are
included in most of the calculations with the perturbative
contribution and condensate contributions. It should be noted that
the direct-instanton approximation is criticized due to
the problem of double counting \cite{Forkel01}, because both
condensate and instanton contributions are included in the
correlator, but most of the gluon condensates could be produced from
instantons.

Up to now, most of the theoretical calculations for the $0^{++}$
scalar glueball were based on QCD Laplace sum rules (LSR), which
emphasize the contributions of the lowest resonance, and have shown
the power in the investigation of the nonperturbative properties of
the hadron ground states. On the other hand, it should be noticed
that the GSR developed later emphasizes only the contribution of the
state considered, and has a cleaner background in comparison with
the LSR. As a cross-check, the same problem should be investigated
using GSR, because both Laplace and Gaussian sum rules are derived
from the same underlying dynamical theory, and should give almost
the same results.

Motivated by the above considerations, our main objective in this
paper is to investigate the $0^{++}$ scalar glueball in the frame
work of Gaussian sum rules. For the correlation function, we include
the contributions from the interactions between the quantum gluons
and the classical instanton background besides the ones comeing
separately from these two different field configurations. For the
spectral function, instead of using the usual zero-width
approximation, we adopt the Breit-Wigner form for the considered
resonance with correct threshold behavior, in order to get the
information of not only the mass scale but also the full decay
width. Moreover, without using the scheme of the mixture of the
traditional condensates and the so-called direct-instanton
contribution, we are working in the framework of the semiclassical
expansion of QCD in the instanton liquid vacuum, a well-defined
self-consistent procedure for the quantum theory justified by the
path-integral quantization formalism. The paper is organized as
follows: In Sec. \ref{sec:CorFunct}, we give the expression for the
correlation function of the $0^{++}$ scalar gluon current. The
spectral function corresponding to this current is constructed in
Sec. \ref{sec:sepctfunct}. Then, a family of the finite-width
Gaussian sum rules are derived in Sec. \ref{sec:FWGSR}. In Sec.
\ref{sec:numerical}, the numerical simulation is carried out.
Finally, a summery of our conclusions and a discussion are given in
Sec. \ref{sec:conclusion}.

\section{Correlation Function}\label{sec:CorFunct}
The correlation function for the scalar glueball in the Euclidean
space-time with a virtuality $q^2$ is defined by
\begin{equation}
\Pi(q^2)=\int d^4x\textrm{e}^{iq\cdot
x}\langle\Omega|TO_s(x)O_s(0)|\Omega\rangle\label{eq:COR},
\end{equation}
where $|\Omega\rangle$ is the physical vacuum, and $O_s$ the scalar
glueball current with the quantum numbers $J^{PC}=0^{++}$
\begin{equation}
O_s=\alpha_sG^a_{\mu\nu}(B)G^{a,\mu\nu}(B)\label{eq:current},
\end{equation}
in which, $\alpha_s$ denotes the strong coupling constant. The
scalar glueball current $O_s$ is gauge-invariant, and
renormalization-invariant at one-loop level. In the spirit of the
semiclassical expansion, and in order to maintain the
$O(4)$ covariance, the gluon field strength tensor $G^a_{\mu\nu}(B)$
is considered as a functional of the full gluon potential, $B_{\mu
a}=A_{\mu a}+a_{\mu a}$, with $A_{\mu a}$ and $a_{\mu a}$ being the
instanton fields and the corresponding quantum fluctuations.

The theoretical expression, $\Pi^{\textrm{QCD}}$, for the
correlation function $\Pi$ may be divided into the following three
parts
\begin{equation}
\Pi^{\textrm{QCD}}(Q^2)=\Pi^{\textrm{pert}}(Q^2)+\Pi^{\textrm{inst}}(Q^2)+\Pi^{\textrm{int}}(Q^2),
  \label{eq:PQCD}
\end{equation}
where $Q^2=q^2$, and $\Pi^{\textrm{pert}}(Q^2)$,
$\Pi^{\textrm{inst}}(Q^2)$, and $\Pi^{\textrm{int}}(Q^2)$ are the
contributions from the only perturbative QCD, the pure instanton
dynamics, and the interactions between the instantons and the
quantum gluon fields, respectively.

The perturbative contribution $\Pi^{\textrm{pert}}(Q^2)$ is already
known to be
\begin{equation}
\Pi^{\textrm{pert}}(Q^2)=Q^4\ln\left(\frac{Q^2}{\mu^2}\right)
\left[a_0+a_1\ln\left(\frac{Q^2}{\mu^2}\right)
+a_2\ln^2\left(\frac{Q^2}{\mu^2}\right)\right]\label{eq:pert},
\end{equation}
where $\mu^2$ is the renormalization scale in the
$\overline{\mbox{MS}}$-dimensional regularization scheme, and the
coefficients with the inclusion of the threshold effects are
\begin{eqnarray}
a_0&=&-2\left(\frac{\alpha_s}{\pi}\right)^2\left[1+\frac{659}{36}
\left(\frac{\alpha_s}{\pi}\right)+247.48\left(\frac{\alpha_s}{\pi}\right)^2\right]
,\nonumber\\
a_1&=&2\left(\frac{\alpha_s}{\pi}\right)^3\left[\frac9{4}
+65.781\left(\frac{\alpha_s}{\pi}\right)\right],\label{eq:pert3}\\
a_2&=&-10.1252\left(\frac{\alpha_s}{\pi}\right)^4.\nonumber
\end{eqnarray}
for QCD with three quark flavors up to three-loop level in the chiral
limit \cite{HSE01,HaS01,CKS97}, and
\begin{eqnarray}
a_0&=&-2\left(\frac{\alpha_s}{\pi}\right)^2\left[1+\frac{51}{4}
\left(\frac{\alpha_s}{\pi}\right)\right]
,\nonumber\\
a_1&=&\frac{11}{2}\left(\frac{\alpha_s}{\pi}\right)^3, \quad
a_2=0.\label{eq:pert2}
\end{eqnarray}
for quarkless QCD up to two-loop level \cite{Bagan90}. Both
expressions for $\Pi^{\textrm{pert}}(Q^2)$ with and without quark
loop corrections are used in our calculation for comparison. With
the assumption that the dominant contribution to
$\Pi^{\textrm{inst}}(Q^2)$ comes from a Belavin-Polyakov-Schwartz-Tyupkin single instanton and
anti-instanton solutions \cite{tHft76,CDG78,BPST75} and the
multi-instanton effects are negligible (see a QCD spectral sum rule (QSSR) approach \cite{Sch95}) , and in view of
the gauge-invariance of the correlation function, one may choose to
work in the regular gauge of the classical single instanton
potential
\begin{eqnarray}
 A^a_{\mu}=\frac{2}{g_s}\eta^a_{\mu \nu}\frac{(x-x_0)_{\nu}}{(x-x_0)^2+\rho^2},
\end{eqnarray}
where $\eta^a_{\mu \nu}$ is the 't Hooft symbol, and $x_0$ and
$\rho$ denote the position and size of the instanton, respectively.
The pure instanton contribution is obtained to be
\cite{NSVZ80,Forkel01,Forkel05,Shu82,Ioffe00}
\begin{equation}
\Pi^{\textrm{inst}}(Q^2)=2^5\pi^2\bar{n}\bar{\rho}^4Q^4K^2_2(\sqrt{Q^2}\bar{\rho})\label{eq:inst},
\end{equation}
where $K_2(x)$ is the McDonald function, $\bar{n}=\int_0^\infty
d\rho n(\rho)$ and $\bar{\rho}$ are the overall instanton density
and the average instanton size in the random instanton background,
respectively.

It is noticed that the contribution to $\Pi^{\textrm{QCD}}$ from the
interactions between instantons and the quantum gluon fields is of
the order of the product of $\alpha_s$ and the overall instanton
density $\bar{n}$. There is no reason to get rid of this
contribution in comparison with the perturbative contributions of
the higher order $\alpha_s^4$ considered in $\Pi^{\textrm{pert}}$.
To calculate such contribution, our key observation is that the
instanton potential $A^a_{\mu}$ obeys also the fixed-point gauge
condition
\begin{equation}
 (x-x_0)_{\mu}A^a_{\mu}(x-x_0)=0
\end{equation}
due to the antisymmetry of the 't Hooft symbols. As a
consequence, the instanton potential can be expressed in terms of
the corresponding field strength tensor as follows:
\begin{equation}
 A^a_{\mu}(x-x_0)=\int_0^1duuF^a_{\mu \nu}[u(x-x_0)](x-x_0)_{\nu}
   \label{eq:gauge-condition}
\end{equation}
and the gauge link with respect to the instanton fields is just the
unit operator, and thus the trace of any product of the
gauge-covariant instanton field strengths at different points is
gauge-invariant. This allows us to conclude that the remainder
quantum corrections to the gauge-invariant correlation function,
arising from the interactions between the instantons and the quantum
gluons, is gauge-invariant as well. Therefore, one may choose any
specific gauge in evaluating the quantum correction to
$\Pi^{\textrm{int}}$.

It is noticed that the interference between the pure quantum part of
the scalar glueball current, $(O_s)_{\textrm{quantum}}$, and the
pure classical instanton part, $(O_s)_{\textrm{instanton}}$ is
vanishing due to the fact that the momentum integration of the
massless gluon propagator or the product of massless quark and gluon
propagators has no scale parameters, and is exactly zero in the
dimensional regularization scheme. The first nonvanishing
contribution comes from the contraction between the two quantum
glueball currents with instanton legs. Working in Feynman gauge, our
result for $\Pi^{\textrm{int}}$ is
\begin{eqnarray}
\Pi^{\textrm{int}}(Q^2)=C_0\alpha_s\bar{n}\pi+\alpha_s^2\bar{n}[C_1
+C_2(Q\bar{\rho})^2\ln(Q\bar{\rho})^2
+C_3\ln(Q\bar{\rho})^2+C_4(Q\bar{\rho})^{-2}]  \label{eq:inte},
\end{eqnarray}
where the coefficients are
\begin{equation}
 C_0=62.62,\,\,  C_1=1533.15,\,\, C_2=825.81,\,\, C_3=-496.33,\,\,
 C_4=-348.89.
\end{equation}
It is remarkable to note that the fixed point $x_0$, which
characterizes the gauge condition (\ref{eq:gauge-condition}),
disappears in the expression of $\Pi^{\textrm{int}}$, as expected from
the gauge invariance of our procedure. The detail calculation for
$\Pi^{\textrm{int}}$ is much involved, and will appear elsewhere.

\section{Spectral function} \label{sec:sepctfunct}
Now, we turn to specify the spectral function for the correlation
function of the scalar glueball current. The imaginary part of the
correlation function Eq. (\ref{eq:PQCD}) is
\begin{eqnarray}
\textrm{Im}\Pi^{\textrm{QCD}}(s)&=&-\pi
s^2\left[a_0+2a_1\ln \frac{s}{\mu^2}+\left(3\ln^2 \frac{s}{\mu^2}-\pi^2\right)a_2\right]\nonumber\\
&&-16\pi^4s^2\bar{n}\bar{\rho}^4J_2(\bar{\rho}\sqrt{s})Y_2(\bar{\rho}\sqrt{s})
+\alpha_s^2\bar{n}\pi(C_2\bar{\rho}^2s-C_3)  \label{eq:ImQCD}.
\end{eqnarray}
The usual lowest resonance plus a continuum model is used to
saturate the phenomenological spectral function,
\begin{equation}
\textrm{Im}\Pi^{\textrm{PHE}}(s)=\rho^{\textrm{had}}(s)
+\theta(s-s_0)\textrm{Im}\Pi^{\textrm{QCD}}(s)
  \label{eq:ImPHE},
\end{equation}
where $s_0$ is the QCD-hadron duality threshold, and
$\rho^{\textrm{had}}(s)$ the spectral function for the lowest scalar
glueball state. Instead of using the zero-width approximation as
usual, the Breit-Wigner form for a single resonance is adopted for
$\rho^{\textrm{had}}(s)$ in the quarkless world
\begin{equation}
\rho^{\textrm{had}}(s)=\frac{f^6m\Gamma}{(s-m^2+\Gamma^2/4)^2+m^2\Gamma^2}
  \label{eq:had},
\end{equation}
where $f^3=\langle\Omega|O_s|0^{++}\rangle$ is the coupling of the
lowest resonance to the scalar glueball current Eq.
(\ref{eq:current}). Recall the threshold behavior for
$\rho^{\textrm{had}}(s)$
\begin{equation}
f^3\rightarrow \lambda_0s, \hspace{20pt}\mbox{for } s\rightarrow 0,
   \label{eq:threshold}
\end{equation}
which is deduced from a low-energy theorem \cite{NSVZ81,Shifman81,Novikov80}. The
early QCD sum rule approach had often used $f^3\rightarrow
\lambda_0s$ in the whole lowest resonance region, however the
obtained mass scale is too low to be expected in comparison with the
lattice QCD results. In fact, the threshold behavior
(\ref{eq:threshold}) is only proven to be valid in the chiral limit;
it may not be extrapolated far away. Therefore, instead of
considering the coupling $f$ as a constant \cite{Sch95}, we choose a
model for $f$ as
\begin{equation}
 f^3=
 \left\{
 \begin{array}{ll}
  \lambda_0s,        &\mbox{for } s<m^2_{\pi}\\
  \lambda_0m^2_{\pi}+\lambda^3,&\mbox{for } s\geq m^2_{\pi}
 \end{array}\right.,
   \label{eq:lambda}
\end{equation}
where $\lambda_0$ and $\lambda$ are some constants determined late
in numerical simulation, so that the spectral function
$\rho^{\textrm{had}}(s)$ has the almost complete Breit-Wigner form
with correct threshold behavior, and in cooperation with the
threshold behavior which is important for the convergence of the
corresponding integral. In Eq. (\ref{eq:lambda}), the constant
$\lambda$ is invoked for the discontinuity at the chiral symmetry
breaking.

Although glueballs are well defined in quarkless QCD, the mixing
with mesons makes the phenomenological side more complicated in full
QCD; all scalar hadron states having a glue content should be
interpolated by the gluonic current, and then present in the
correlator with different couplings. In Refs.
\cite{Cheng06,Vicente06,Frank05,Amsler08,Amsler95,Amsler195}, the
glueball is shared between the three scalar hadrons. These three
hadrons are very close in mass; the assumption of single resonance
maybe be not appropriate. The form of the spectral function for
three resonances we will use is taken to be
\begin{equation}
\rho^{\textrm{had}}(s)=\sum_{i=1}^{3}\frac{f_i^6m_i\Gamma_{i}}{(s-m_i^2
+\Gamma^2_{i}/4)^2+m_{i}^2\Gamma_{i}^2}
  \label{eq:had2},
\end{equation}
where all $f_i$ have the form shown in (\ref{eq:lambda}) with the
same $\lambda_0$, because $\lambda_0$ is fixed to be 5 GeV by the
low-energy theorem of QCD, and so that its value is independent of
what an individual resonance considered.

\section{Finite-width Gaussian sum rules}\label{sec:FWGSR}
A family of Gaussian sum rules can also be constructed from the
Borel transformation of the correlation function in Eq. (\ref{eq:PQCD})
\cite{Bertlmann85}
\begin{equation}
\mathcal{G}^{\textrm{had}}_{k}(s_0;\hat{s},\tau)
=\mathcal{G}^{\textrm{QCD}}_{k}(s_0;\hat{s},\tau)
+\frac1{\sqrt{4\pi\tau}}\exp\left[-\frac{\hat{s}^2}{4\tau}\right]\Pi(0)\delta_{k,-1},
   \label{eq:GSR}
\end{equation}
where $\Pi(0)$ comes from the subtraction to the corresponding
dispersion relation due to the degree of divergence of the
correlation function of the scalar glueball, and
\begin{eqnarray}
\mathcal{G}^{\textrm{QCD}}_k(s_0;\hat{s},\tau)&=&\mathcal{G}^{\textrm{QCD}}_k(\hat{s},\tau)-\mathcal{G}^{\textrm{cont}}_{k}(s_0;\hat{s},\tau),\\
 \mathcal{G}^{\textrm{had}}_{k}(s_0;\hat{s},\tau)
 &=&\frac1{\sqrt{4\pi\tau}}\int^{s_0}_0dss^k
 \exp\left[-\frac{(s-\hat{s})^2}{4\tau}\right]\frac1{\pi}\rho^{\textrm{had}}(s),
\end{eqnarray}
where $\mathcal{G}^{\textrm{cont}}_{k}(s_0;\hat{s},\tau)$ is the
contribution of continuum,
\begin{equation}
 \mathcal{G}^{\textrm{cont}}_{k}(s_0;\hat{s},\tau)\equiv
 \frac1{\sqrt{4\pi\tau}}\int^{\infty}_{s_0}dss^k
     \exp\left[-\frac{(s-\hat{s})^2}{4\tau}\right]\frac1{\pi}\mathrm{Im}\Pi^{\textrm{QCD}}(s)
\end{equation}
and $\mathcal{G}^{\textrm{QCD}}_k(\hat{s},\tau)$ is defined  as
\begin{equation}
\mathcal{G}^{\textrm{QCD}}_k(\hat{s},\tau)\equiv\frac{2\tau}{\sqrt{4\pi\tau}}\hat{\mathcal{B}}
    \left[\mbox{Im}\frac{(\hat{s}+iQ^2)^k\Pi^{\textrm{QCD}}(-\hat{s}-iQ^2)}{Q^2}\right],
\end{equation}
with the Borel transformation $\hat{\mathcal{B}}$ being defined by
\begin{equation}
\hat{\mathcal{B}}\equiv{\lim_{\left.\begin{subarray}{c}N\rightarrow\infty\\Q^4
\rightarrow\infty\end{subarray}\right|}}_{Q^4/N\equiv4\tau}
\frac{(-1)^N}{(N-1)!}\,(Q^4)^N\left(\frac{\textrm{d}}{\textrm{d}Q^4}\right)^N
\label{eq:BorTG}.
\end{equation}
The Gaussian sum rule emphasizes only the contribution of the hadron
state considered, and suppresses the background exponentially
(according to the Gaussian distribution). Recall that the Laplace
sum rule stress only the contribution from the lowest state, and
suppresses the other contributions exponentially (according to the
exponential distribution).

For $k=-1$, $0$, and $1$, a straightforward but tedious manipulation
leads to
\begin{eqnarray}
&&\mathcal{G}^{\textrm{QCD}}_{-1}(\hat{s},\tau)\nonumber\\
&=&-\frac1{\sqrt{4\pi\tau}}\int^{\infty}_{0}dss\exp\left[-\frac{(s-\hat{s})^2}{4\tau}\right]\nonumber\\
&&\times\left[(a_0-\pi^2a_2)+2a_1\ln\left(\frac{s}{\sqrt{\tau}}\right)+3a_2\ln^2\left(\frac{s}{\sqrt{\tau}}\right)\right]\nonumber\\
&&-\frac{1}{\sqrt{4\pi\tau}}\int^{\infty}_{0}dss\exp\left[-\frac{(s-\hat{s})^2}{4\tau}\right]16\pi^3\bar{n}\bar{\rho}^4J_2\left(\bar{\rho}\sqrt{s}\right)Y_2\left(\bar{\rho}\sqrt{s}\right)\nonumber\\
&&+\frac{1}{\sqrt{4\pi\tau}}\int^{\infty}_{0}ds\exp\left[-\frac{(s-\hat{s})^2}{4\tau}\right]\bar{n}\alpha_s^2(C_2\bar{\rho}^2)\nonumber\\
&&+\frac1{\sqrt{4\pi\tau}}\exp\left[-\frac{\hat{s}^2}{4\tau}\right]\left[C_4\alpha_s^2\bar{n}\bar{\rho}^{-2}\frac{\hat{s}}{4\tau}-(C_0\alpha_s\bar{n}\pi+C_1\alpha_s^2\bar{n}+2^7\pi^2\bar{n})\right]\nonumber\\
&&+\frac1{\sqrt{4\pi\tau}}\exp\left[-\frac{\hat{s}^2}{4\tau}\right]\bar{n}\alpha_s^2C_3
\left\{\frac{3\pi}{12}\left[\textrm{erf}\left(\frac{\hat{s}}{2\sqrt{\tau}}\right)\right]^2+\frac{\pi}2\textrm{erf}\left(\frac{\hat{s}}{2\sqrt{\tau}}\right)
\right.\nonumber\\
&&\left.-\ln\left(\sqrt{\tau}\bar{\rho}^2\right)+\frac{\gamma}2-\ln2\right\},\label{eq:GQCD}
\end{eqnarray}

\begin{eqnarray}
&&\mathcal{G}^{\textrm{QCD}}_{0}(\hat{s},\tau)\nonumber\\
&=&-\frac1{\sqrt{4\pi\tau}}\int^{\infty}_{0}dss^2\exp\left[-\frac{(s-\hat{s})^2}{4\tau}\right]\nonumber\\
&&\times\left[(a_0-\pi^2a_2)+2a_1\ln\left(\frac{s}{\sqrt{\tau}}\right)+3a_2\ln^2\left(\frac{s}{\sqrt{\tau}}\right)\right]\nonumber\\
&&-\frac{1}{\sqrt{4\pi\tau}}\int^{\infty}_{0}dss^2\exp\left[-\frac{(s-\hat{s})^2}{4\tau}\right]16\pi^3\bar{n}\bar{\rho}^4J_2\left(\bar{\rho}\sqrt{s}\right)Y_2\left(\bar{\rho}\sqrt{s}\right)\nonumber\\
&&+\frac{1}{\sqrt{4\pi\tau}}\int^{\infty}_{0}ds\exp\left[-\frac{(s-\hat{s})^2}{4\tau}\right]\bar{n}\alpha_s^2(C_2\bar{\rho}^2s-C_3)\nonumber\\
&&+\frac1{\sqrt{4\pi\tau}}\exp\left[-\frac{\hat{s}^2}{4\tau}\right]C_4\alpha_s^2\bar{n}\bar{\rho}^{-2},\label{eq:GQCD0}
\end{eqnarray}

\begin{eqnarray}
&&\mathcal{G}^{\textrm{QCD}}_{1}(\hat{s},\tau)\nonumber\\
&=&-\frac1{\sqrt{4\pi\tau}}\int^{\infty}_{0}dss^3\exp\left[-\frac{(s-\hat{s})^2}{4\tau}\right]\nonumber\\
&&\times\left[(a_0-\pi^2a_2)+2a_1\ln\left(\frac{s}{\sqrt{\tau}}\right)+3a_2\ln^2\left(\frac{s}{\sqrt{\tau}}\right)\right]\nonumber\\
&&-\frac{1}{\sqrt{4\pi\tau}}\int^{\infty}_{0}dss^3\exp\left[-\frac{(s-\hat{s})^2}{4\tau}\right]16\pi^3\bar{n}\bar{\rho}^4J_2\left(\bar{\rho}\sqrt{s}\right)Y_2\left(\bar{\rho}\sqrt{s}\right)\nonumber\\
&&+\frac{1}{\sqrt{4\pi\tau}}\int^{\infty}_{0}dss\exp\left[-\frac{(s-\hat{s})^2}{4\tau}\right]\bar{n}\alpha_s^2(C_2\bar{\rho}^2s-C_3).\label{eq:GQCD1}
\end{eqnarray}

\section{numerical analysis}\label{sec:numerical}
Now, we specify the input parameters in the numerical calculation.
The color and flavor numbers are taken to be $N_c=3$ and $N_f=3$,
respectively. The expressions for two-loop quarkless ($N_f=0$)
running coupling constant $\alpha_s(Q^2)$ at renormalization scale
$\mu$ \cite{Groom00,Prosperi06}
\begin{equation}
\frac{\alpha^{(2)}_s(\mu^2)}{\pi}=\frac1{\beta_0L}-\frac{\beta_1}{\beta_0}\frac{\ln
L}{(\beta_0L)^2},\label{eq:alpha2}
\end{equation}
and for the three-loop running coupling constant with three flavors
($N_f=3$)
\begin{equation}
\frac{\alpha_s(\mu^2)}{\pi}=\frac{\alpha^{(2)}_s(\mu^2)}{\pi}
+\frac1{(\beta_0L)^3}\left[L_1\left(\frac{\beta_1}{\beta_0}\right)^2+\frac{\beta_2}{\beta_0}\right],\label{eq:alpha3}
\end{equation}
are used, where the central value of the $\overline{\textrm{MS}}$
QCD scale $\Lambda$ is taken to be $120$ MeV, and
\begin{equation}
\begin{array}{l}
L=\ln\left(\frac{\mu^2}{\Lambda^2}\right),\quad L_1=\ln^2L-\ln L-1,\\
\beta_0=\frac1{4}\left[11-\frac2{3}N_f\right],\quad
 \beta_1=\frac1{4^2}\left[102-\frac{38}{3}N_f\right],\\
\beta_2=\frac1{4^3}\left[\frac{2857}{2}-\frac{5033}{18}N_f+\frac{325}{54}N_f\right].
\end{array}
\end{equation}
Recall here that research on the renormalization group improvement
for Gaussian sum rules indicates that $\mu^2=\sqrt{\tau}$
\cite{Narison81}. The subtraction constant $\Pi(0)$ is fixed by QCD
low-energy theorem \cite{NSVZ80}
\begin{equation}
\Pi(0)=\frac{32\pi}{9}\langle\alpha_s
G^2\rangle\simeq0.6\,\textrm{GeV}^4.
\end{equation}
The values of the average instanton size and the overall instanton
density are adopted from the instanton liquid model \cite{Shu82}
\begin{equation}
\bar{n}=1\,\textrm{fm}^{-4}=0.0016\,\textrm{GeV}^4,\,\,\,
\bar{\rho}=\frac1{3}\,\textrm{fm}=1.689\,\textrm{GeV}^{-1}.
\end{equation}
Finally, the mass of the neutral pion is taken from the
experimental data, i.e. $m_{\pi}=135$ MeV.

In order to measure the compatibility between both sides of the sum
rules (\ref{eq:GSR}) realized in our numerical simulation, we
introduce a variation, $\delta$, defined by
\begin{equation}
 \delta=\frac{1}{N}\sum_1^N\frac{[L(\tau_i)-R(\tau_i)]^2}{|L(\tau_i)R(\tau_i)|},
\end{equation}
where the interval $[\tau_{\textrm{min}}, \tau_\textrm{max}]$ is
divided into $100$ equal small intervals, $N=101$, and $L(\tau_i)$
and $R(\tau_i)$ are left-hand side and right-hand side of Eq.(\ref{eq:GSR}) evaluated
at $\tau_i$.

To determine the values of the resonance parameters in Eq.
(\ref{eq:had}), we match both sides of sum rules Eq. (\ref{eq:GSR})
optimally in the fiducial domain. In doing so, the value of
$\hat{s}$ should approximately be set to be the corresponding mass
squared of the resonance $m^2$. To suppress the continuum
contribution, we require $\hat{s}\leq m^2$. The conditions for
determining the value of $s_0$ are: first, it should be grater than
$m^2$; and second, it should guarantee that there exists a sum rule
window for our Gaussian sum rules. We note that the upper limit
$\tau_{\textrm{max}}$ of the sum rule window is determined by
requiring that the contribution from the continuum should be less
than that of the resonance
\begin{equation}
\mathcal{G}^{\textrm{cont}}_{k}(s_0;\hat{s},\tau_{\textrm{max}})
\leq
\mathcal{G}^{\textrm{QCD}}_{k}(s_0;\hat{s},\tau_{\textrm{max}}),\label{eq:upt}
\end{equation}
while the lower limit $\tau_{\textrm{min}}$ of the sum rule window
is obtained by requiring the contribution of pure instantons to be
greater than 50\% of
$\mathcal{G}^{\textrm{QCD}}_k(s_0;\hat{s},\tau)$, because such
classical contributions should be dominant in the low-energy region.
Moreover, to require that the multi-instanton corrections remain
negligible, we simply adopt a rough estimate
\begin{equation}
 \tau_{\textrm{min}}^{-1}\leq (2\bar{\rho})^4\sim
 \left(\frac{2}{0.6\,\textrm{GeV}}\right)^{4}.
\end{equation}
With these requirements, the figurations and numerical results are
given below.

\begin{figure}[!hbp]
\begin{center}
\subfigure[]{\includegraphics[angle=0,width=8.5cm]{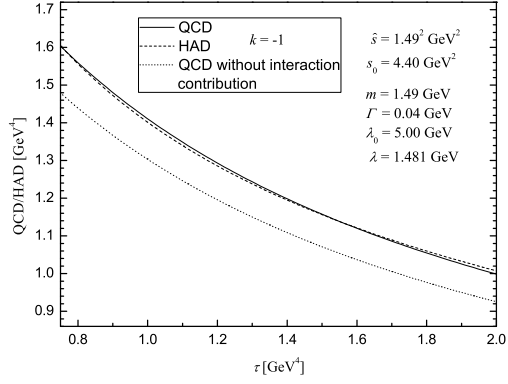}}\vspace{-0.5cm}
\subfigure[]{\includegraphics[angle=0,width=8.5cm]{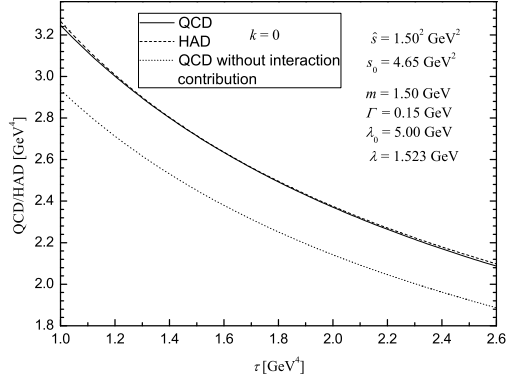}}\vspace{-0.5cm}
\subfigure[]{\includegraphics[angle=0,width=8.5cm]{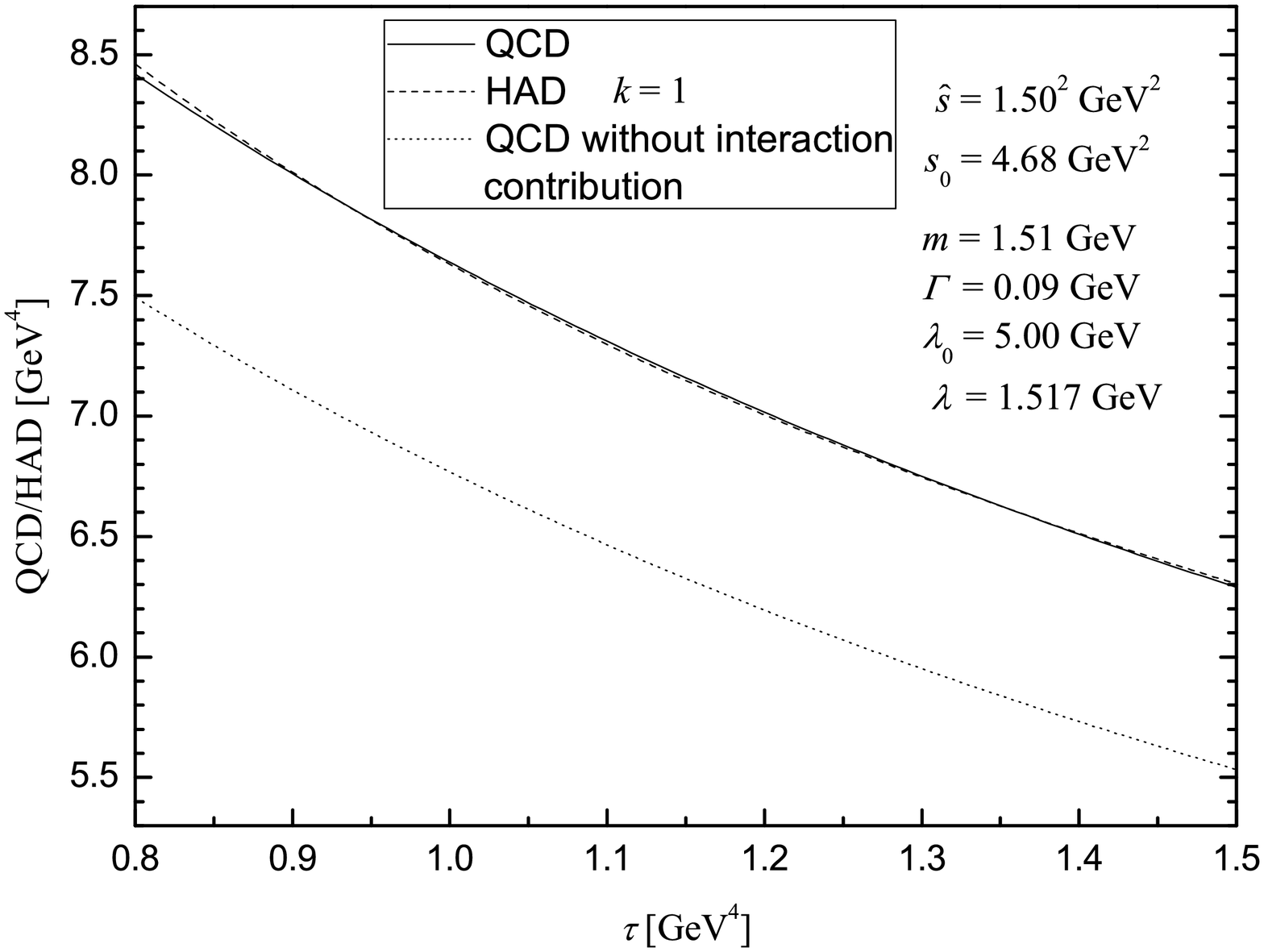}}
\caption{\small The curves for the left-hand side and right-hand side of the Eq.
(\ref{eq:GSR}) for quarkless QCD with only the lowest resonance
considered. The solid line denotes the right-hand side (QCD), dashed line for
left-hand side (HAD), and dotted line for the right-hand side (QCD) without the interaction
contribution of the Gaussian sum rules
(\ref{eq:GSR}).}\label{fig:GBW-GSR-1500}
\end{center}
\end{figure}

For the case of quarkless QCD with the lowest resonance in the
$0^{++}$ channel, we adopt the isolated lowest resonance model
(\ref{eq:had}) for the spectral function, the optimal parameters
governing the sum rules are listed in Table \ref{tab:GBW}.
\begin{table}[!h]
\caption{\small The fitting values of the mass $m$ and width
 $\Gamma$ of the lowest $0^{++}$ scalar glueball, and of the parameters $\lambda$ and $f$,
 $s_0$, $[\tau_{\textrm{min}},\tau_{\textrm{max}}]$, and $\delta$
 characterizing the couplings to the lowest resonance, the continuum threshold, the sum rule window,
 and the compatibility measure for finite-width Gaussian sum rules (\ref{eq:GSR}) of $k=-1$, $0$, and $1$
 in quarkless QCD for a given $\hat{s}$, the real part of the complex
 $q^2$ ($q^2=\hat{s}+iQ^2$).}
\begin{center}
\begin{footnotesize}
\begin{tabular}{crccccccc}\hline\hline
$\hat{s}$($\textrm{GeV}^2$)& $k$& $m$(GeV)& $\Gamma$(GeV)&
$\lambda$(GeV)& $f$(GeV)& $s_0$($\textrm{GeV}^2$)&
$[\tau_{\textrm{min}},\tau_{\textrm{max}}](\textrm{GeV}^4)$& $\delta$\\
\hline
$1.49^2$& $-1$& 1.49& 0.04& 1.481& 1.495& 4.40&  [0.75-2.0]& $2.40\times10^{-5}$\\
$1.50^2$&    0& 1.50& 0.15& 1.523& 1.534& 4.65&  [1.0-2.6]&  $7.79\times10^{-6}$\\
$1.50^2$&    1& 1.51& 0.09& 1.517& 1.530& 4.68&  [0.8-1.5]&  $2.80\times10^{-6}$\\
\hline\hline \label{tab:GBW}
\end{tabular}
\end{footnotesize}
\end{center}
\end{table}
The corresponding curves for the left-hand side and right-hand side of (\ref{eq:GSR})
of $k=-1$, $0$, and $+1$ are displayed in Fig. \ref{fig:GBW-GSR-1500}
where the solid lines are the right-hand side (QCD) of Eq. (\ref{eq:GSR}), and
the dashed lines are the left-hand side (HAD) of Eq. (\ref{eq:GSR}), and
the dotted lines are for the right-hand side (QCD) excluding the contribution of
interactions between the instantons and the quantum gluons (the same
for hereafter).
Taking the average, the values of the mass and width of the lowest
$0^{++}$ scalar glueball living in quarkless QCD, and the
corresponding optical fit parameters are predicted to be
\begin{eqnarray}
 &&m=1.52\pm0.18\,\textrm{GeV},\, \Gamma=0.2\pm 0.15\,\textrm{GeV}, \nonumber\\
 &&f=1.47\pm0.13\,\textrm{GeV},\,s_0=3.8\pm0.9\,\textrm{GeV}^2.
    \label{eq:average1}
\end{eqnarray}
where the errors are estimated from the uncertainties of the spread
between the individual sum rules, and by varying the
phenomenological parameters, $\Lambda$ and
$\langle\alpha_sG^2\rangle$, appropriately away from their central
values (the same for hereafter)
\begin{eqnarray}
 \Lambda&=&120-200\,\textrm{MeV},   \\
 \langle\alpha_sG^2\rangle&=&0.6-0.8\,\textrm{GeV}^4.
\end{eqnarray}

\begin{figure}[!hbp]
\begin{center}
\subfigure[]{\includegraphics[angle=0,width=8.5cm]{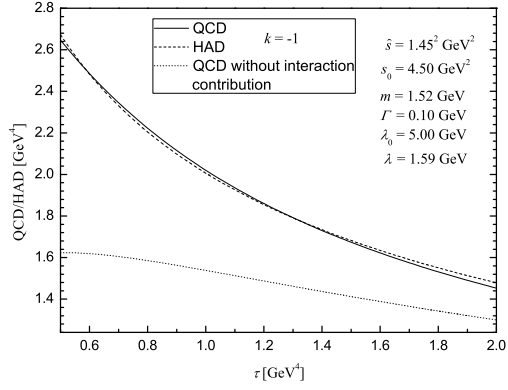}}\vspace{-0.5cm}
\subfigure[]{\includegraphics[angle=0,width=8.5cm]{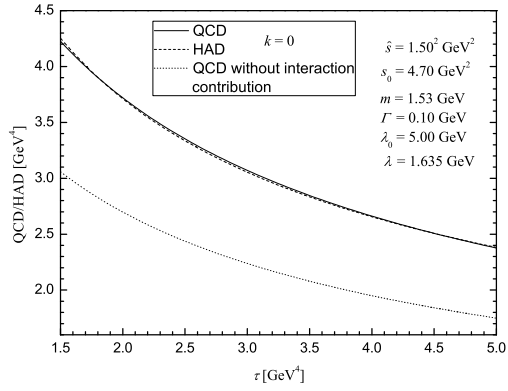}}\vspace{-0.5cm}
\subfigure[]{\includegraphics[angle=0,width=8.5cm]{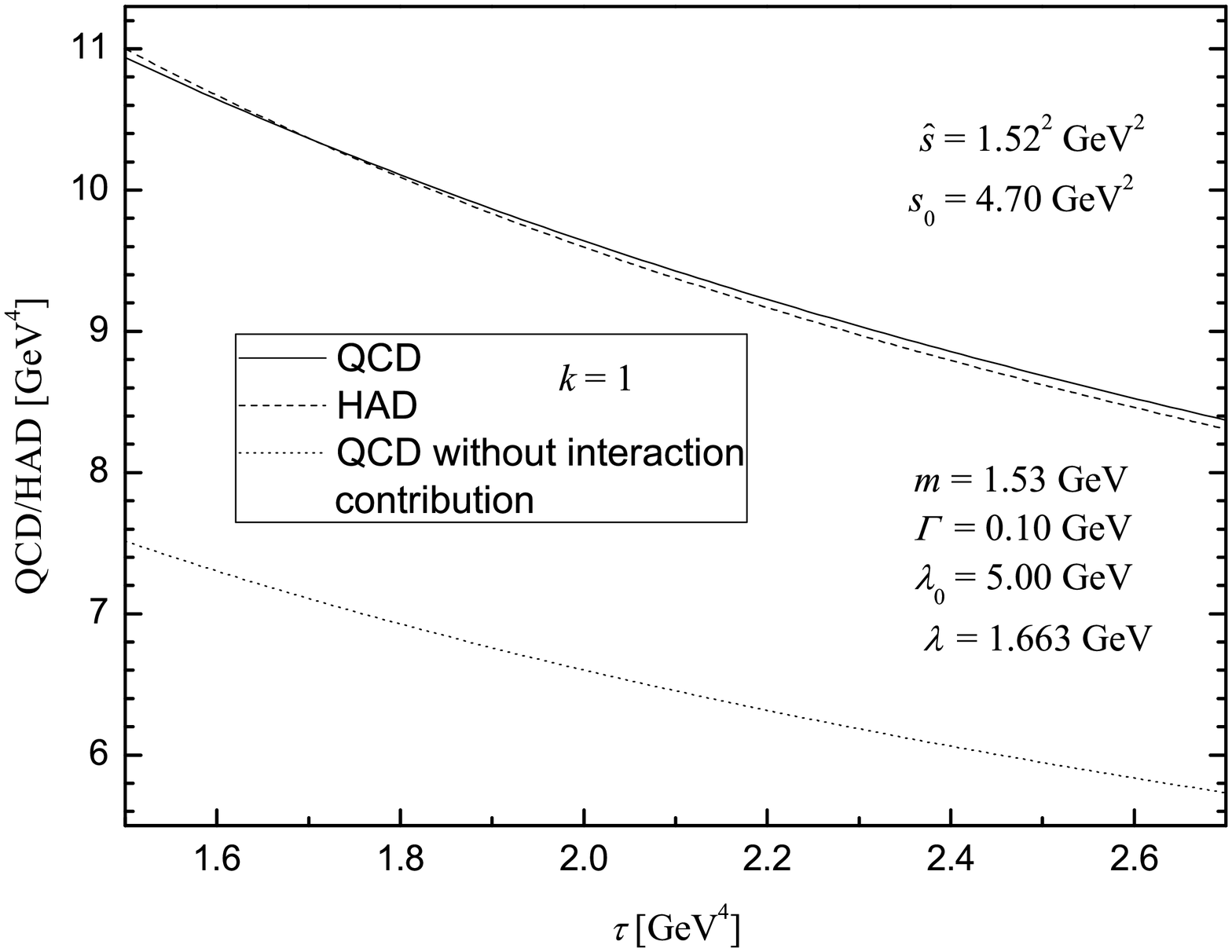}}
\caption{\small The curves for the left-hand side and right-hand side of the Eq.
(\ref{eq:GSR}) for three-flavor QCD in chiral limit with only the
lowest resonance considered. The solid line denotes the right-hand side (QCD),
dashed line for left-hand side (HAD), and dotted line for the right-hand side (QCD) without
the interaction contribution of the Gaussian sum rules
(\ref{eq:GSR}).} \label{fig:BWR-GSR-1500}
\end{center}
\end{figure}

For quarkless QCD, there is only one well-defined scalar bound state
of gluon suggested by lattice QCD and also from our investigation
just described above. Including quarks enhances the difficulty of
the task since many states possessing the same quantum numbers may
be present in the correlator. Even so, at the first step as
comparison, we still consider only the lowest resonance in the
$0^{++}$ channel for the case of QCD with three massless quarks as
usual for the scalar glueball mass $m$ of about 600 MeV, because that
resonance $f_0(600)$ may be considered to be well isolated. The
optimal parameters governing the sum rules are listed in
Table \ref{tab:BWR}.
\begin{table}[!h]
\caption{\small The fitting values of the mass $m$ and width
 $\Gamma$ of the lowest $0^{++}$ scalar glueball, and of the parameters $\lambda$ and $f$,
 $s_0$, $[\tau_{\textrm{min}},\tau_{\textrm{max}}]$, and $\delta$
 characterizing the couplings to the lowest single resonance, the continuum threshold, the sum rule window,
 and the compatibility measure for finite-width Gaussian sum rules (\ref{eq:GSR}) of $k=-1$, $0$, and $1$
 in QCD with three massless quarks for a given $\hat{s}$, the real part of the complex
 $q^2$ ($q^2=\hat{s}+iQ^2$).}
\begin{center}
\begin{footnotesize}
\begin{tabular}{crccccccc}\hline\hline
$\hat{s}$($\textrm{GeV}^2$)& $k$& $m$(GeV)& $\Gamma$(GeV)&
$\lambda$(GeV)& $f$(GeV)& $s_0$($\textrm{GeV}^2$)&
$[\tau_{\textrm{min}},\tau_{\textrm{max}}](\textrm{GeV}^4)$& $\delta$\\
\hline
$1.45^2$& $-1$& 1.52& 0.10& 1.590& 1.602& 4.5& [0.5-2.0]& $6.87\times10^{-5}$\\
$1.50^2$&  0& 1.53& 0.10& 1.635& 1.646& 4.7&  [1.5-5.0]& $1.34\times10^{-5}$\\
$1.52^2$&  1& 1.53& 0.10& 1.663& 1.674& 4.7&  [1.5-2.7]& $3.04\times10^{-5}$\\
\hline\hline \label{tab:BWR}
\end{tabular}
\end{footnotesize}
\end{center}
\end{table}
The corresponding curves for the left-hand side and right-hand side of (\ref{eq:GSR})
of $k=-1$, $0$ and, $+1$ are displayed in Fig.\ref{fig:BWR-GSR-1500}.
Taking the average, the values of the mass and width of the
(probable isolated) lowest $0^{++}$ scalar glueball in the world of
QCD with three massless quarks, and the corresponding optical fit
parameters are predicted to be
\begin{eqnarray}
 &&m=1.54\pm0.17\,\textrm{GeV},\, \Gamma=0.23\pm 0.13\,\textrm{GeV}, \nonumber\\
 &&f=1.64\pm0.14\,\textrm{GeV},\,s_0=3.8\pm0.9\,\textrm{GeV}^2.
    \label{eq:average2}
\end{eqnarray}

\begin{figure}[!hbp]
\begin{center}
\subfigure[]{\includegraphics[angle=0,width=8.5cm]{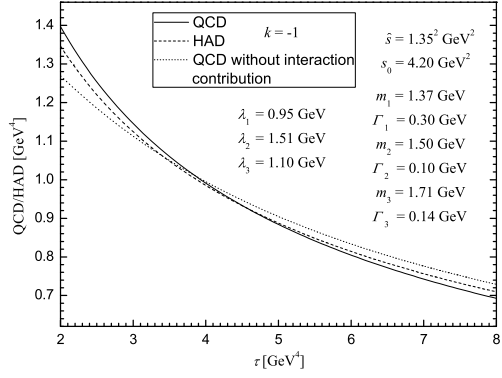}}\vspace{-0.5cm}
\subfigure[]{\includegraphics[angle=0,width=8.5cm]{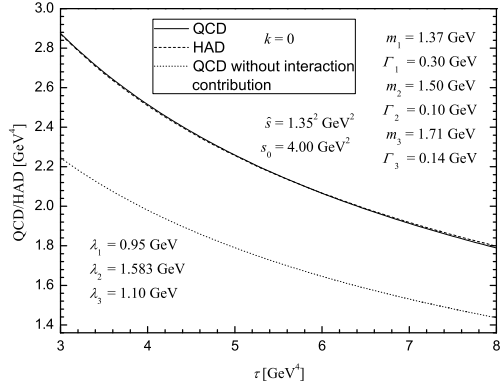}}\vspace{-0.5cm}
\subfigure[]{\includegraphics[angle=0,width=8.5cm]{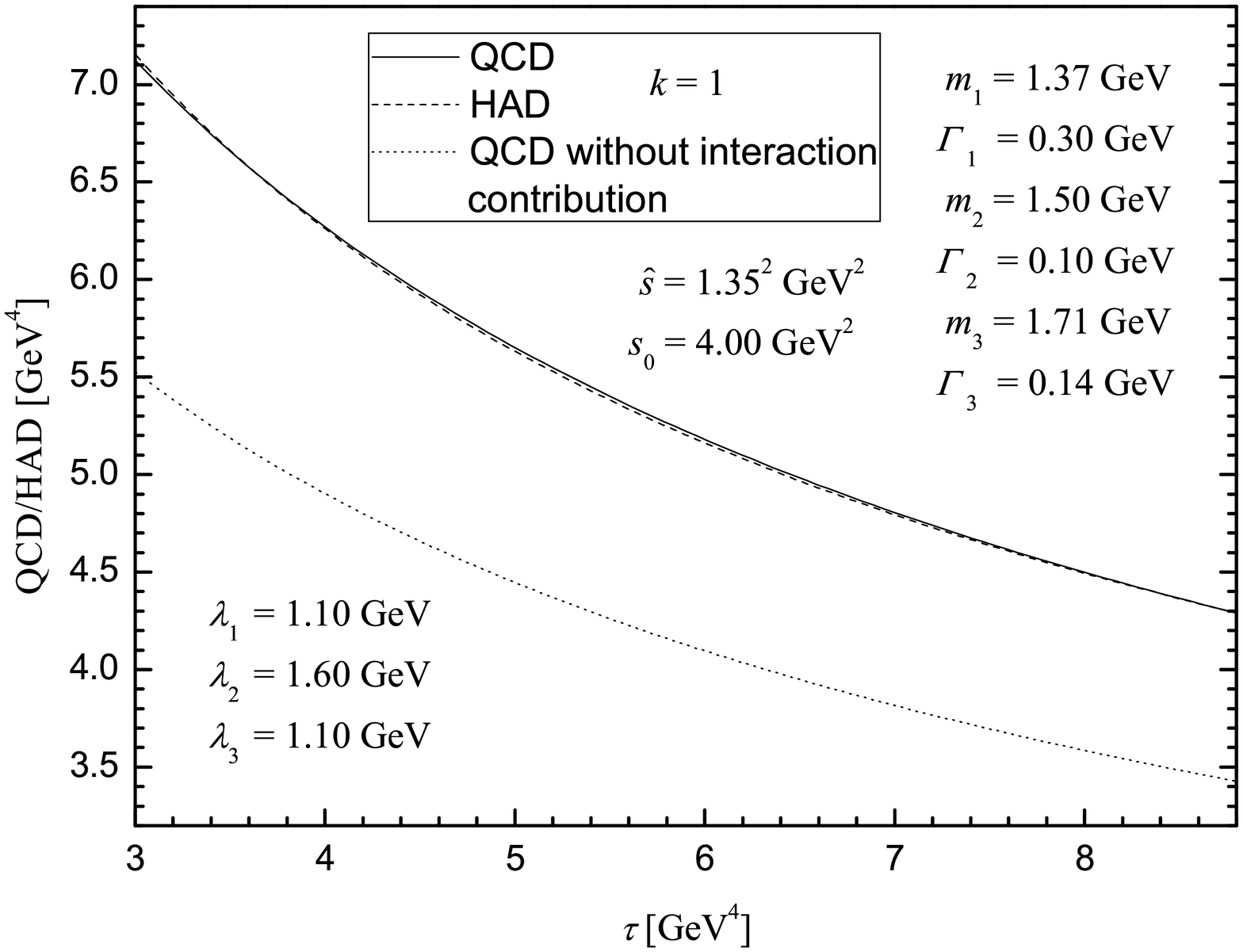}}
\caption{\small The curves for the left-hand side and right-hand side of the Eq.
(\ref{eq:GSR}) for three-flavor QCD in chiral limit with the lowest
three resonances considered. The solid line denotes the right-hand side (QCD),
dashed line for left-hand side (HAD), and dotted line for the right-hand side (QCD) without
the interaction contribution of the Gaussian sum rules
(\ref{eq:GSR}).}\label{fig:3BW-GSR-1370}
\end{center}
\end{figure}

The above one isolated lowest resonance assumption is questioned
from the admixture with quarkonium states, and from the experimental
data that three $0^{++}$ scalar states are around the mass scale of
$1500$ MeV [namely $f_0(1370)$, $f_0(1500)$, and $f_0(1710)$].
Therefore, we adopt the three-resonance model for the
phenomenological side of the sum rules for the case of QCD with
three massless quarks in the $0^{++}$-channel when $m$ is above 1 GeV;
the optimal parameters governing the sum rules are listed in
Table \ref{tab:3BWR}.
\begin{table}[!h]
 \caption{\small The fitting values of the mass $m$ and width
 $\Gamma$ of the lowest $0^{++}$ scalar glueball, and of the parameters $\lambda$ and $f$,
 $s_0$, $[\tau_{\textrm{min}},\tau_{\textrm{max}}]$, and $\delta$
 characterizing the couplings to the three closely located resonances,
 the continuum threshold, the sum rule window,
 and the compatibility measure for finite-width Gaussian sum rules (\ref{eq:GSR}) of $k=-1$, $0$, and $1$
 in QCD with three massless quarks for a given $\hat{s}$, the real part of the complex
 $q^2$ ($q^2=\hat{s}+iQ^2$).}
\begin{center}
\begin{footnotesize}
\begin{tabular}{crccccccc}\hline\hline
$\hat{s}$($\textrm{GeV}^2$)& $k$&  $m$ (GeV)& $\Gamma$(GeV)&
$\lambda$(GeV)& $f$(GeV)& $s_0$($\textrm{GeV}^2$)&
$[\tau_{\textrm{min}},\tau_{\textrm{max}}](\textrm{GeV}^4)$& $\delta$\\
\hline
& & 1.37& 0.30& 0.950& 0.983& & \\
$1.35^2$& $-1$& 1.50& 0.10& 1.510& 1.523& 4.2& [2.0,8.0]& $2.90\times10^{-4}$\\
& & 1.71& 0.14& 1.100& 1.125& & \\
\hline
& & 1.37& 0.30& 0.950& 0.983& & \\
$1.35^2$& 0& 1.50& 0.10& 1.583& 1.595& 4.0&  [3.0,8.0]& $6.42\times10^{-6}$\\
& & 1.71& 0.14& 1.100& 1.125& & \\
\hline
& & 1.37& 0.30& 1.100& 1.125& & \\
$1.35^2$& 1& 1.50& 0.10& 1.600& 1.612& 4.0&  [3.0,8.8]& $5.13\times10^{-6}$\\
& & 1.71& 0.14& 1.100& 1.125& & \\
\hline\hline \label{tab:3BWR}
\end{tabular}
\end{footnotesize}
\end{center}
\end{table}
The corresponding curves for the left-hand side and right-hand side of (\ref{eq:GSR})
of $k=-1$, $0$, and $+1$ are displayed in Fig.
\ref{fig:3BW-GSR-1370}.
Taking the average, the values of the widths of the three lowest
$0^{++}$ scalar resonances in the world of QCD with three massless
quarks, and the corresponding optical fit parameters are predicted
to be
\begin{eqnarray}
m=1.37\pm0.06\,\textrm{GeV},\, \Gamma=0.30\pm 0.10\,\textrm{GeV},\,
f=1.10\pm0.13\,\textrm{GeV}.
    \label{eq:average3}
\end{eqnarray}
for the resonance $f_0(1370)$, and
\begin{eqnarray}
m=1.50\pm0.10\,\textrm{GeV},\, \Gamma=0.10\pm 0.06\,\textrm{GeV}\,
f=1.60\pm0.11\,\textrm{GeV}.
    \label{eq:average4}
\end{eqnarray}
for $f_0(1500)$, and
\begin{eqnarray}
 &&m=1.71\pm0.11\,\textrm{GeV},\, \Gamma=0.14\pm 0.08\,\textrm{GeV}\,f=1.10\pm0.14\,\textrm{GeV}.
    \label{eq:average5}
\end{eqnarray}
for $f_0(1710)$.

Figures.\ref{fig:GBW-GSR-1500}-\ref{fig:3BW-GSR-1370} show the
consistent match between the both sides of Eq. (\ref{eq:GSR}) for
$k=-1,0$, and $1$, respectively, with the fitting parameters. The
matching between both sides of the sum rules is very well over the
whole fiducial region with a very little departure.

These results are in good accordance with the experimental data of
$f_0(1500)$, $m=1505\pm5$ MeV, $\Gamma=109\pm7$ MeV \cite{Amsler08},
and the sum rule calculation of Ref. \cite{Forkel01}, $m=1.53\pm0.2$
GeV, $f=1.01\pm0.25$ GeV (see Table \ref{tab:Numer}). We do not
calculate the higher moments sum rule, because for $k>1$, the
continuum contributions become very large.
\begin{table}[!h]
 \caption{\small The numerical results from other methods. }
\begin{center}
\begin{footnotesize}
\begin{tabular}{b{1.5cm}b{2.5cm}m{2.5cm}p{3cm}p{2cm}p{1.7cm}}\hline\hline
Methods& Mass (GeV)& Width (GeV)& Coupling (GeV)& $s_0\,(\textrm{GeV}^2)$& References\\
\hline GSR& 0.8 - 1.6& 0.4 - 0.6& & 2.3& \cite{HaS01}\\
\hline QLQCD & 1.3 - 1.7& 0.035 - 0.873& & &
\cite{Meng09,Chen06,Colin99,Vaccar99,Sexton95}\\
\hline QSSR & $1.4\pm0.2$& & $2.56-2.61$& & \cite{Sch95}\\
\hline& $1.25\pm0.2$& & $1.05\pm0.1$& $5.0\pm0.1$&
\cite{Forkel05}\\
LSR & $1.53\pm0.2$& & $1.01\pm0.25$& $5.0\pm0.1$&
\cite{Forkel01}\\
& $1.52\pm0.2$& & $0.39\pm0.145$& $4.2\pm0.2$&
\cite{Narison98}\\
\hline & $1.42-1.5$& & & & \cite{Frank05}\\
model & 1.666& & & &\cite{Cheng06}\\
& 1.633& & & & \cite{Aniso97}\\
\hline\hline \label{tab:Numer}
\end{tabular}
\end{footnotesize}
\end{center}
\end{table}

\section{DISCUSSION AND CONCLUSIONS}\label{sec:conclusion}
The properties of the $0^{++}$ scalar glueball are examined in a family
of the finite-width Gaussian sum rules. The correlation function is
calculated in a semiclassical expansion, a well-defined process
justified in the path-integral quantization formalism, of QCD in the
instanton background, namely the instanton liquid model of the QCD
vacuum. Besides the contributions from pure gluons and instantons
separately, the one arising from the interactions between the
classical instanton fields and the quantum gluon ones are taken into
account as well. Instead of using the usual zero-width approximation
for the spectral function of the considered resonances, the
Breit-Wigner form for the resonances with a correct threshold
behavior is adopted. With the QCD standard input parameters, three
Gaussian sum rules with the $k=-1,0$, and 1-th moments are carefully
studied.

For the quarkless QCD, we have,in fact, changed the value of
$\hat{s}$, and found that the value of the mass of the lowest
resonance is approximately proportional to $\hat{s}$, and the value
of $s_0$ arrives almost at its maximum for $\hat{s}$ lying between
1.50 GeV$^2$ and 1.70 GeV$^2$, where the couplings to the state are
almost the same for different $k$, and the corresponding widths
become small and stable. We have only shown the situation with
the optimal compatibility. We note here that the value of $\delta$
for $m=650$ MeV is one or two-orders lower than the optimal one, and
the values of $f$ are not coincident for different $k$, so that the
mass scale of the lowest $0^{++}$ scalar glueball may not be lower
than 1 GeV. The mass and width of the lowest glueball without quark
loop corrections are predicted in Eq. (\ref{eq:average1}).

For QCD with three massless flavors and by considering only single
scalar resonance, the same behavior with respect to the changing of
$\hat{s}$ appears. Namely, the value of the mass of the resonance is
approximately proportional to $\hat{s}$, and the value of $s_0$
arrives almost at its maximum for $\hat{s}$ lying between
1.35 GeV$^2$ and 1.70 GeV$^2$, where the couplings to the state are
almost the same for different $k$, and the corresponding widths
become small and stable. When $\hat{s}=1.70^2$ GeV$^2$ (the
situation with optimal compatibility), all physical parameters are
almost the same for different $k$. We note here that the value of
$\delta$ for $m=650$ MeV is nearly one order lower than the optimal one,
and the values of $f$ are not coincident for different $k$ as in the
case of pure QCD, so that the mass scale of the lowest $0^{++}$
scalar glueball, even in the world of QCD with massless quarks, may
not be lower than 1 GeV. The mass and width of the lowest glueball
with quark loop corrections under the assumption of no mixture
between glueball and $q\bar{q}$ state are predicted in Eq.
(\ref{eq:average2}).

For QCD with three massless flavors and by considering three closely
located $0^{++}$ scalar resonances $f_0(1370)$, $f_0(1500)$, and
$f_0(1710)$ (namely both $\hat{s}$ and the resonance masses are
given input parameters), the behavior with respect to the changing
of $\hat{s}$ is changed. Namely, the values of the widths and $s_0$
for the three resonances remain as almost invariant, and only
their couplings slightly increase when increasing $\hat{s}$. The
optimal compatibility is arrived at $\hat{s}=1.35^2$ GeV$^2$, as
shown in Table \ref{tab:3BWR} and Fig.\ref{fig:3BW-GSR-1370}. The
widths of the lowest three $0^{++}$ resonances coupled to the
glueball current $O_s$ are predicted in (\ref{eq:average3}), from
which we can read off the corresponding couplings $f^3$ to the three
resonances $f_0(1370)$, $f_0(1500)$, and $f_0(1710)$ with masses $1.35$,
$1.47$, and $1.70$ GeV as
\begin{equation}
 0.95\pm 0.47\,\textrm{GeV}^3,\,\,\,
 3.92\pm 0.85\,\textrm{GeV}^3,\,\,\,
 1.42\pm 0.51\,\textrm{GeV}^3,
  \label{eq:3couplings}
\end{equation}
respectively.

In summary, we may conclude that, first, any $0^{++}$ scalar
resonance below 1 GeV, such as $f_0(600)$, contains almost no component of the scalar glueball;
second, the values of the mass and decay width of the $0^{++}$
resonance, in which the fraction of the scalar glueball state is
dominant, are $m=1.50\pm0.10$ GeV and $\Gamma=0.10\pm 0.06$ GeV,
respectively, and the value of its coupling to the corresponding
current is $f^3_{s\geq m^2_{\pi}}=3.92\pm 0.85\ \,\textrm{GeV}^3$;
third, the fractions of the scalar glueball contained in the other
nearby $0^{++}$ scalar resonances, $f_0(1370)$ and $f_0(1710)$, are
also appreciable. These are not only compatible with lattice QCD
simulation \cite{Meng09,Chen06,Colin99,Vaccar99,Sexton95} and other
estimations \cite{Forkel05, Forkel01, Sch95, Narison98, Liang93}, but
also in good accordance with the experimental data of the low scalar
resonances \cite{Amsler08,Amsler95,Amsler195}.

It is also remarkable that the three Gaussian sum rules lead to
almost the same results, a consistency between the subtracted and
unsubtracted sum rules is very well justified. We note that we have
not been working within the mixed scheme, namely with including
condensates, and in the same time, adopting the so-called direct-instanton approximation, but simply with a self-consistent
framework, a quantum theory in a classical background, without the
problem of double counting. In this aspect, our results further
justified the instanton liquid model for QCD among many other
justifications.

In our semiclassical expansion, the leading contribution to the sum
rules comes from instantons themselves, especially in the region
below the threshold $s_0$. It is the amount of this contribution that
determines the low bound of the sum rule window. This means that
the nonlinear configurations of gluons have a dominant role with
respect to the quantum fluctuations in the low-energy region.

The contribution of the interactions between the classical instanton
fields and quantum gluon ones, considered in this paper but
neglected in earlier sum rule calculations
\cite{Forkel05,Forkel01,HSE01,HaS01,Narison98}, is in fact not
negligible. To the contrary, its amount is approximately double or
even triple that from the pure quantum fluctuations in the whole
fiducial domain, expected from a view point of the semiclassical
expansion. Moreover, it is obviously seen from Figs.
\ref{fig:GBW-GSR-1500}-\ref{fig:3BW-GSR-1370} that, without taking
the contribution from the interactions between instantons and
quantum gluons into account, the departures between
$\mathcal{G}^{\textrm{had}}_{k}(s_0;\hat{s},\tau)$ and
$\mathcal{G}^{\textrm{QCD}}_{k}(s_0;\hat{s},\tau)$  without
interaction become large, and all three
Gaussian sum rules become less stable, and thus less reliable.

Finally, it should be noticed that the imaginary part of the instanton
contribution is an oscillating, amplifying and nonpositive
defined function, and so is the imaginary part of the correlation
function. This property which is a fatal problem for the QCD sum
rule calculation with the instanton background, may make the
contribution of continuum too large to be under control. Hilmar
Forkel introduced a Gaussian distribution for the instanton to get rid
off this trouble, and obtained a smaller $0^{++}$ mass scale:
$1.25\pm0.2$ GeV \cite{Forkel05} compared to the earlier result
$1.53\pm0.2$ GeV \cite{Forkel01}. We did not use this Gaussian
distribution, but simply chose a smaller fitting parameter to avoid
this problem.

\begin{acknowledgments}
We are grateful to Prof. H. G. Dosch for useful discussions. This
work is supported by the National Natural Science Foundation of
China under Grant No. 10075036, BEPC National Laboratory Project
R\&D and BES Collaboration Research Foundation,Chinese Academy of
Sciences (CAS) Large-Scale Scientific Facility Program.
\end{acknowledgments}


\end{document}